\documentclass[twocolumn,epsfig,graphics,noshowpacs,floatfix,nofootinbib]{revtex4-1}

\usepackage{amsmath,amsfonts,amssymb,graphics,graphicx,epsfig,color,times,bbm}
\usepackage{amsthm}
\usepackage{psfrag}
\usepackage{braket}
\usepackage{xcolor}
\usepackage[normalem]{ulem}

\usepackage{chngcntr}
\counterwithin{paragraph}{section}

\AtBeginDocument{\usepackage{booktabs}}
\graphicspath{{figures/}}
\usepackage[hidelinks]{hyperref}
\hypersetup{colorlinks=false}
\usepackage{array}
\newcolumntype{P}[1]{>{\centering\arraybackslash}p{#1}}
\newcolumntype{M}[1]{>{\centering\arraybackslash}m{#1}}
\usepackage{graphicx}

\usepackage[framemethod=tikz]{mdframed}
\usepackage{hyperref}

\definecolor{warning_bgcol}{RGB}{252,248,229}
\definecolor{warning_textcol}{RGB}{111,89,54}
\definecolor{warning_linecol}{RGB}{252,248,229}
\definecolor{danger_bgcol}{RGB}{239,223,222}
\definecolor{danger_textcol}{RGB}{128,60,57}
\definecolor{danger_linecol}{RGB}{239,223,222}
\definecolor{success_bgcol}{RGB}{224,237,216}
\definecolor{success_textcol}{RGB}{68,104,60}
\definecolor{success_linecol}{RGB}{224,237,216}
\definecolor{info_bgcol}{RGB}{220,237,246}
\definecolor{info_textcol}{RGB}{58,100,126}
\definecolor{info_linecol}{RGB}{220,237,246}

\mdfdefinestyle{box_style}{
  skipabove=.7\baselineskip,
  skipbelow=.7\baselineskip,
  innertopmargin=.65\baselineskip,
  innerbottommargin=.65\baselineskip,
  innerleftmargin=.5\baselineskip,
  innerrightmargin=.5\baselineskip,
  splittopskip=1.5\baselineskip,
  splitbottomskip=\baselineskip,
  roundcorner=.3\baselineskip
}
\mdfdefinestyle{warning_style}{
  style=box_style,
  backgroundcolor=warning_bgcol,
  linecolor=warning_linecol,
  fontcolor=warning_textcol,
}
\mdfdefinestyle{success_style}{
  style=box_style,
  backgroundcolor=success_bgcol,
  linecolor=success_linecol,
  fontcolor=success_textcol,
}
\mdfdefinestyle{danger_style}{
  style=box_style,
  backgroundcolor=danger_bgcol,
  linecolor=danger_linecol,
  fontcolor=danger_textcol,
}
\mdfdefinestyle{info_style}{
  style=box_style,
  backgroundcolor=info_bgcol,
  linecolor=info_linecol,
  fontcolor=info_textcol,
}

\newtheorem{definition}{Definition}

\begin{document}

\title{Increasing error tolerance in quantum computers with dynamic bias arrangement}

\author{Hector Bomb\'in}
\author{Chris Dawson}
\author{Naomi Nickerson}
\author{Mihir Pant}\email{Lead author: mihir@psiquantum.com}
\author{Jordan Sullivan}

\affiliation{PsiQuantum, Palo Alto}
\date\today

\begin{abstract}
Many quantum operations are expected to exhibit bias in the structure of their errors. A fixed bias can be exploited~\cite{bonilla2021xzzx, claes2022tailored, sahay2022tailoring, sahay2023high} to improve error tolerance by statically arranging the errors in beneficial configurations. In some cases an error bias can be dynamically reconfigurable, an example being linear optical fusion where the basis of a fusion failure can be chosen before the measurement is made. Here we introduce methods for increasing error tolerance in this setting by using classical decision-making to adaptively choose the bias in measurements as a fault tolerance protocol proceeds. We study this technique in the setting of linear optical fusion based quantum computing (FBQC). We provide examples demonstrating that by dynamically arranging erasures, the loss tolerance can be tripled when compared to a static arrangement of biased errors while using the same quantum resources: we show that for the best FBQC architecture of~\cite{bartolucci2021fusion} the threshold increases from $2.7\%$ to $7.5\%$ per photon with the same resource state by using dynamic biasing. Our method does not require any specific code structure beyond having a syndrome graph representation. We have chosen to illustrate these techniques using an architecture which is otherwise identical to that in~\cite{bartolucci2021fusion}, but deployed together with other techniques, such as different fusion networks, higher loss thresholds are possible.
\end{abstract}

\maketitle

\section{Introduction}

Increasing error tolerance is an important objective in the path to fault tolerant quantum computation. As hardware develops and quantum architectures become more mature we are gaining a greater understanding of the structure of physical noise. This structure can be exploited to improve tolerance to errors~\cite{aliferis2008fault,tuckett2018ultrahigh,aliferis2009fault}. Several recent works have considered methods of exploiting bias in the noise which is a commonly arising error structure in many quantum computing platforms~\cite{bonilla2021xzzx, claes2022tailored, sahay2022tailoring, sahay2023high}. These works are based on the underlying technique of dimensionality reduction of the errors, for example within a 3-dimensional topological fault tolerance protocol the noisy measurements are confined to 2-dimensional planes which reduces the probability of logical error. We refer to these techniques as `static bias arrangement'. While these methods have led to impressive gains in error thresholds, they depend on the errors being biased by an order of magnitude or more.  The error biases are chosen in a deterministic manner, forming static lower dimensional objects and errors linking these planes quickly wash away the advantage.

However in many physical platforms, including photonics, it may not be necessary to choose the configuration of biases in advance in a deterministic pattern. Instead, they may be chosen dynamically, at the time of the operation. In this paper, we present ``dynamic bias arrangement", a method for increasing error tolerance by choosing the bias in operations based on previous measurement outcomes. We show that dynamic bias arrangement can increase error tolerance significantly compared to static arrangement of errors. We present two examples from a photonic fusion-based quantum computation (FBQC) architecture~\cite{bartolucci2021fusion} where this technique triples loss tolerance: from $0.48\%$ to $1.6\%$ when using six qubit resource states, and from $2.7\%$ to $7.5\%$ with Shor-encoded six qubit resource states.

Our dynamic bias arrangement method does not depend on any specific structure of the fault tolerance protocol, with the decision of how to bias a measurement being based only on graph properties related to the outcomes of neighboring measurements using techniques inspired by explosive percolation theory~\cite{achlioptas2009explosive}. We also find that dynamic bias arrangement is able to operate with more moderately biased measurements in comparison to static bias arrangement methods (section~\ref{subsec:encoded_fusion}).

\begin{figure}[ht!]
    \centering
    \includegraphics[width=\columnwidth]{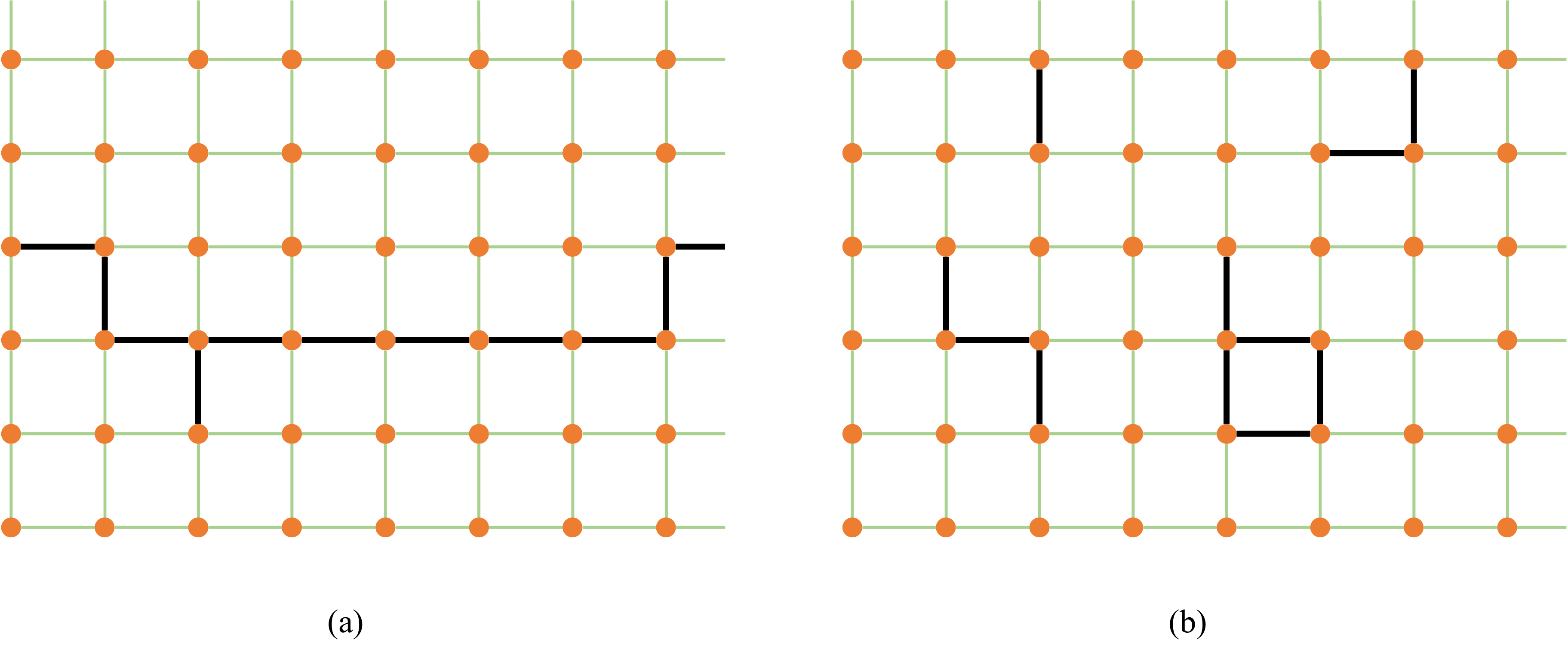}
    \caption{The arrangement of erasures determines whether a fault tolerant protocol suffers a logical erasure or not: two examples of syndrome graphs are shown with the same number of erasures where (a) has a logical erasure and (b) does not have logical erasure. The syndrome graph is a compact representation of the error structure where vertices represent checks, edges represent potential error locations, and here bold edges represent the locations at which there was an erasure. A logical erasure occurs when a path of erased edges spans the lattice. The syndrome graph is introduced in Section~\ref{subsec:FBQC_recap}.}
    \label{fig:logical_erasure_vs_not}
\end{figure}

The central idea of dynamic bias arrangement is that not all errors in a fault-tolerant quantum computer are equally damaging, and given some information about past events we can choose to distribute errors in less damaging locations. This is illustrated in the two syndrome graphs depicted in Fig.~\ref{fig:logical_erasure_vs_not}(a) and (b) which both have the same number of erasures. The erasure pattern in Fig.~\ref{fig:logical_erasure_vs_not}(a) forms a long chain that crosses the syndrome graph that is indicative of a logical erasure. In comparison, the erasure pattern in Fig.~\ref{fig:logical_erasure_vs_not}(b) consists of isolated erasures which is much more benign, and does not cause a logical error.  

In order to make choices about how to distribute errors we use techniques similar to decoding to analyze errors that have already occurred and compute a score for each upcoming operation. Based on that information we then decide which variation of the operation to make in the next time step. The result is that the error configuration becomes far from i.i.d despite the error probabilities remaining unchanged. In doing this the tolerance to errors can be significantly increased.

Dynamic bias arrangement is particularly well suited for photonic quantum computation with dual rail qubits where the fundamental operations are fusions. The erasure in fusion measurements is strongly biased, and can be chosen in one of two configurations using single qubit rotations prior to measurement ~\cite{rudolph2017optimistic, bartolucci2021fusion, bartolucci2021switch}. Furthermore, here we exploit the feature that interleaving-based photonic architectures~\cite{bombin2021interleaving} involve a  sequential series of operations, which allows extra information to be used for each bias decision. The methods we introduce here are designed and simulated for a fusion-based quantum computing architecture~\cite{bartolucci2021fusion} in a photonic system. However the techniques we introduce are general purpose and can be applied in any context where there is a dynamically reconfigurable bias in errors.

\section{Fusion-based quantum computation and Linear optics}

\subsection{Recap: Fusion-based quantum computation}
\label{subsec:FBQC_recap}

In this paper, we will study techniques for bias handling in the setting of fusion-based quantum computation (FBQC)~\cite{bartolucci2021fusion}. FBQC is a model for fault tolerant quantum computing that is naturally suited to the physical primitives readily accessible in photonic systems. These are entangling measurements, called \textit{fusions}, which are performed on the qubits of small constant sized entangled \textit{resource states}. A network of fusions acting on resource states is called a \textit{fusion network}.

\begin{figure*}
\includegraphics[width =0.9\textwidth]{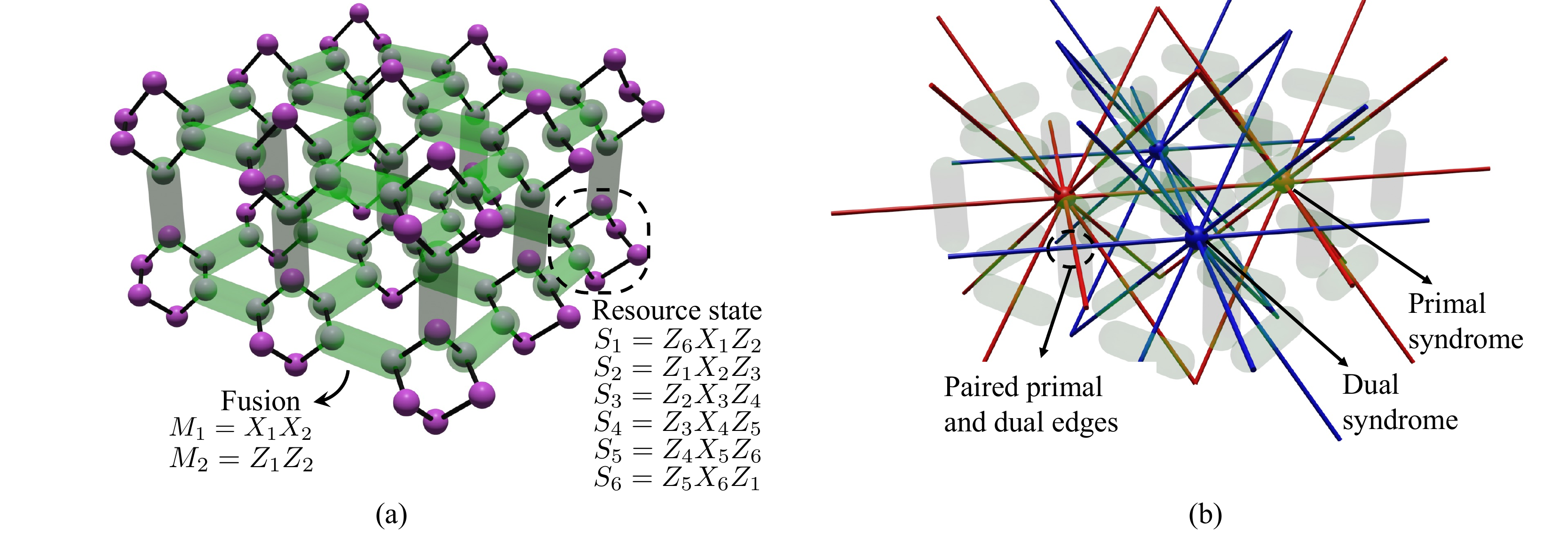}
\caption{The ``6-ring" fusion network. 
(a) Layout of resource states and fusions. Each resource state is a graph state in the form of a ring (black) of 6 qubits (pink). Two qubit fusions are shown in green. All fusion measurements in the fusion network are two qubit projective measurements projective measurements on the bases $M_1 = X_1X_2$ and $M_2 = Z_1Z_2$.
(b) Syndrome graph resulting from the fusion network. Primal edges and syndromes (spheres) are in red and dual edges and syndromes are in blue. Each fusion contributes one measurement/edge in the primal and one in the dual. Both the primal and dual syndrome graphs have an equal number of edges from $XX$ and $ZZ$ measurement outcomes.
}
\centering
\label{fig:kagome6_layout}
\end{figure*}

We will consider fusion-based quantum computation schemes derived from the surface code, and use schemes where resource states and fusions can both be described using the stabilizer formalism. Fig.~\ref{fig:kagome6_layout}(a) shows an example of a fusion network called the ``6-ring" fusion  network that can be used for fault-tolerant quantum computation. The resource states for this fusion network are graph states~\cite{hein2004multiparty} in the form of rings of six qubits (qubits are in pink and the edges of the graph state are in black). The fusions (green capsules) measure the stabilizers $X_1\otimes X_2$ and $Z_1\otimes Z_2$ on the input qubits, also known as a ``Bell fusion". In the rest of the paper, we will remove the tensor product symbol while writing operators for brevity e.g. we will write the measurement stabilizers above as $X_1X_2$ and $Z_1Z_2$.

The root of fault-tolerance in FBQC is redundancy between resource state stabilizers and measurement stabilizers. The measurement outcomes and their redundancy can be graphically represented with a \textit{syndrome graph}. Every edge in a syndrome graph represents a measurement and every vertex represents a parity check. Fig.~\ref{fig:kagome6_layout}(b) shows the syndrome graph for the 6-ring fusion network. Every measurement in the bulk is part of two parity checks. In fusion networks derived from surface codes, including the 6-ring fusion network, the syndrome graph locally splits into two disconnected pieces. We call these the ``primal" and ``dual" syndrome graphs, shown in red and blue respectively in Fig.~\ref{fig:kagome6_layout}(b). In the 6-ring fusion network, every fusion produces two measurement results corresponding to the measurement of the $X_1X_2$ and $Z_1Z_2$ stabilizers. One of these measurement results corresponds to an edge in the primal syndrome graph and the other corresponds to an edge in the dual syndrome graph as depicted by the ``paired primal and dual edges" in Fig.~\ref{fig:kagome6_layout}(b). Further details on fusion based quantum computation can be found in \cite{bartolucci2021fusion}. 

The 6-ring fusion network will be used as an example to illustrate the concepts in this paper. However, the method can also be used to improve the error tolerance of other fusion networks, and beyond FBQC the techniques can be applied in other models of computation where there is dynamically reconfigurable bias in errors. The syndrome graph formulation provides a straightforward way to map these methods to other architectures and models of computation.

\subsection{Linear optical fusion}

\begin{figure}[ht!]
    \centering
    \includegraphics[width=\columnwidth]{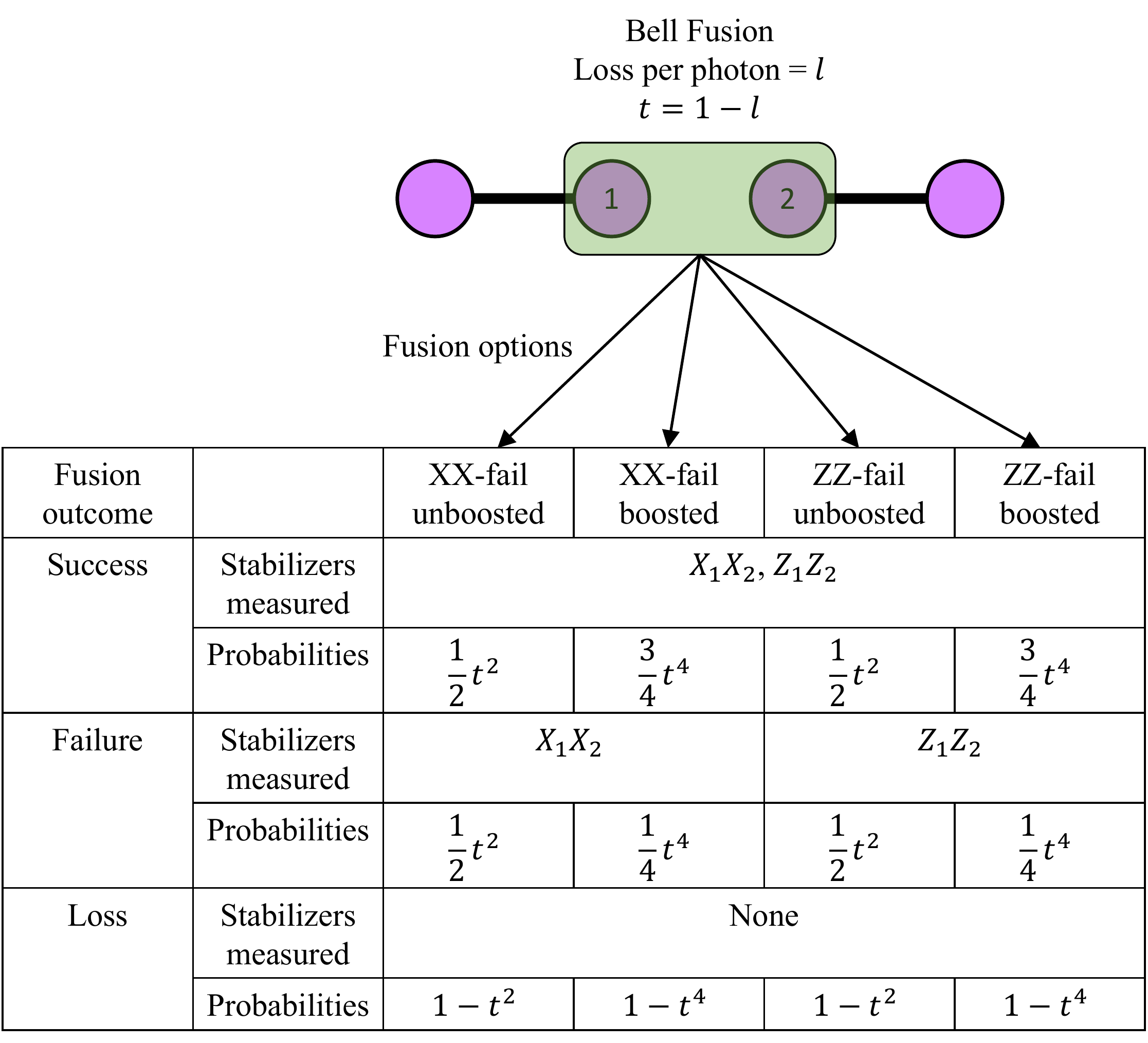}
    \caption{
    Outcomes of a linear optical Bell fusion. A fusion on qubits from two separate cluster states is shown, with intended outcomes $X_1 X_2$ and $Z_1 Z_2$. In the presence of photon loss there are three possible outcomes: \emph{fusion success} where both measurement outcomes are obtained, \emph{fusion failure} where only one of the outcomes is obtained, and \emph{fusion loss} where no measurement outcome is obtained. Fusion failure is intrinsic to linear optics, and can happen even when all operations are ideal. Fusion loss only occurs due to errors in the system, most commonly if one or more of the photons going into the fusion measurement are lost. }
    \label{fig:fusion_outcomes}
\end{figure}

We now describe the behavior of Bell fusion with dual rail qubits and linear optical operations, which is the physical operation that will be our source of reconfigurable biased errors.
Bell fusion on dual-rail qubits can be implemented using a linear optical circuit in which all four modes of the two qubits are measured. 
This is often referred to as type-II fusion~\cite{browne2005resource}. 
This Bell fusion is probabilistic and are three types of outcomes, which are illustrated in Figure~\ref{fig:fusion_outcomes}: 

\begin{enumerate}
    \item {\bf Success.} Fusion ``succeeds" with probability $1-p_{\text{fail}}$, measuring the input qubits in the Bell stabilizer basis $ X_1X_2, Z_1Z_2$ as intended\footnote{assuming that the two qubits are from separate states initially. A fusion on two qubits that are already correlated can potentially succeed with unit probability}. 
    \item {\bf Failure.} The fusion ``fails" with probability $p_{\text{fail}}$, in which case it performs separable single qubit measurements of either $Z_1I_2, I_1Z_2$ or $X_1I_2, I_1X_2$, depending on the type of circuit used to perform the fusion. This freedom of choosing which measurements are performed on failure is the source of bias arrangement. 
    \item {\bf Loss.} In the presence of photon loss or other imperfections then there is a third possible outcome: fusion ``loss". If a photon is lost, or if a multiphoton error occurs, the fusion mesaurement may detect the wrong number of photons. In this case neither intended stabilizer outcome is measured. The loss probability, $l$, depends on the number of photons involved in the measurement.
\end{enumerate}

Linear optical fusion has a highly biased error model. But there is another very important characteristic: there are different variations of how the fusion can be performed which change its bias. 
In the original form of Bell fusion~\cite{browne2005resource}, when the fusion fails, $X_1X_2$ (inferred from measurement of $X_1I$ and $IX_2$) is measured and $Z_1Z_2$ erased. We refer to this as an ``$XX$-fail fusion". However, if a pair of Hadamards were performed on both qubits prior to the fusion, the status of $X$ and $Z$ would be flipped i.e. success would measure the same stabilizers $X_1X_2$ and $Z_1Z_2$, failure would measure only $Z_1Z_2$ (inferred from measurement of $Z_1I$ and $IZ_2$) and again fusion loss would result in no measurement. As a result, with the Hadamards, $X_1X_2$ has a higher probability of being erased than $Z_1Z_2$. We call this a ``$ZZ$-fail fusion". We call the Pauli operator measured in case of fusion failure the ``failure basis" of the fusion.

The nature of this reconfigurable fusion operation is similar to that of ``bias preserving gates" in which gates which differ by a Hadamard in their action can be implemented without affecting the basis of the bias of errors introduced~\cite{puri2020bias,lescanne2020exponential,xu2022engineering,claes2022tailored}. This freedom of picking which measurement to prioritize based on previous fusion outcomes is the key feature of fusion which is exploited to improve error tolerance in this paper;
by deciding which measurement is erased in the case of failure, we can prioritize one measurement over the other. 

In addition to changing the failure basis, we also have the option of ``boosting" fusion. The original Bell fusion~\cite{browne2005resource} circuit has only two qubit inputs (the two qubits being measured) and $p_{\text{fail}} = 1/2$. We call this ``unboosted" fusion. Subsequently, it was shown that by adding an additional Bell pair as input to the circuit, it is possible to achieve $p_{\text{fail}} = 1/4$. Boosting reduces the probability of fusion failure but increases the probability of fusion loss which can be more detrimental. In the regime of loss per photon less than $1-\sqrt{6/7} \approx 7.4\%$, boosting can be used to reduce the sum of erasures in the primal and dual syndrome graphs but it increases the probability of losing both fusion outcomes simultaneously. Therefore, boosting is less desirable in the high loss regime and in fusions where one fusion outcome is significantly more important than the other.

Changing the failure basis does not change the overall rate of erasures across the primal and dual syndrome graphs. However, as we will show, the judicious choice of failure bases allows us to arrange erasures in the syndrome graphs in the form more like that shown in Fig.\ref{fig:logical_erasure_vs_not}(b) and avoid configurations of the form shown in Fig.\ref{fig:logical_erasure_vs_not}(a).

\subsection{Error model}
\label{subsec:error_model}

We consider an error model that accounts for both photon loss and Pauli error (measurement error). 

\textbf{Photon loss.}
Each photon in the fusion is independently lost with probability $l$. If no photon in a fusion is lost, it succeeds with probability $1-p_{\text{fail}}$ and fails with probability $p_{\text{fail}}$.  
We define the transmission probability $t = 1-l$. For unboosted fusion, $p_{\text{fail}} = 1/2$, and with probability $1-t^2$ at least one photon in the fusion is lost, such that the fusion produces the loss outcome where both measurements are erased. Therefore, for unboosted fusion, the probability of success is $t^2/2$ and the probability of failure is also $t^2/2$. If fusion is boosted, the failure probability is reduced to $p_{\text{fail}} = 1/4$. However, boosting requires an additional Bell pair in the fusion. To account for this, we assume that the photons in the Bell pair used in boosting also see the same loss $l$. As a result the probability of the loss outcome is $1-t^4$. The probability of success is $3t^4/4$ and the probability of failure is $t^4/4$. The outcomes of fusion under different choices and their probabilities are summarized in Fig.~\ref{fig:fusion_outcomes}.

\textbf{Measurement error.}
We assume that each fusion measurement outcome has a probability $p_m$ of suffering a bit flip, such that it returns an incorrect parity outcome. All measurement errors are treated as independent.
Fusion measurement errors capture the effect of both Pauli-X errors and Pauli-Z errors on resource state qubits. A physical X (or Z) error on a resource state can lead to either a primal or dual error, depending on its location in the fusion network. These Pauli errors can arise from a multitude of mechanisms including photon distinguishability and imperfections in the physical components performing the fusion.

Comparing to a phenomenological model of the surface code, these fusion measurement errors behave like both Pauli-X and Pauli-Z errors \textit{and} stabilizer measurement errors. In this analogy, errors that affect fusion outcomes that correspond to edges in the primal syndrome graph can be thought of as (phenomenological) Pauli-X errors or plaquette measurement errors. Errors that affect fusion outcomes corresponding edges in the dual syndrome graph can be thought of as (phenomenological) Pauli-Z errors, or errors in star-type stabilizer measurements.

\section{Static bias arrangement}
\label{sec:SBA}

We first present a method for arranging erasures statically, i.e. without any information from previous measurements, when there is a bias in the measurements. We specifically consider the fusion error model introduced in the previous section where some of the measurement outcomes have higher erasure rates than others. Approaches using such static bias arrangement methods have been used in previous work in the literature~\cite{bonilla2021xzzx, claes2022tailored, sahay2022tailoring}. We will illustrate how this technique can be used in FBQC and use these results as the baseline to explore dynamic bias arrangement in the next section.

For the 6-ring fusion network from Fig.~\ref{fig:kagome6_layout}, the primal and dual syndrome graphs have the same structure. Both have an erasure threshold of $11.98\%$~\cite{bartolucci2021fusion}. Hence, if the fusion failure basis is chosen randomly, the fusion network cannot tolerate fusion failure with a probability of more than $2 \times 11.98\% = 23.96 \%$ even in the absence of loss and measurement error. Hence, even boosted fusion is insufficient for quantum computation if the fusion basis is chosen randomly. 

\begin{figure}[ht!]
    \centering
    \includegraphics[width=\columnwidth]{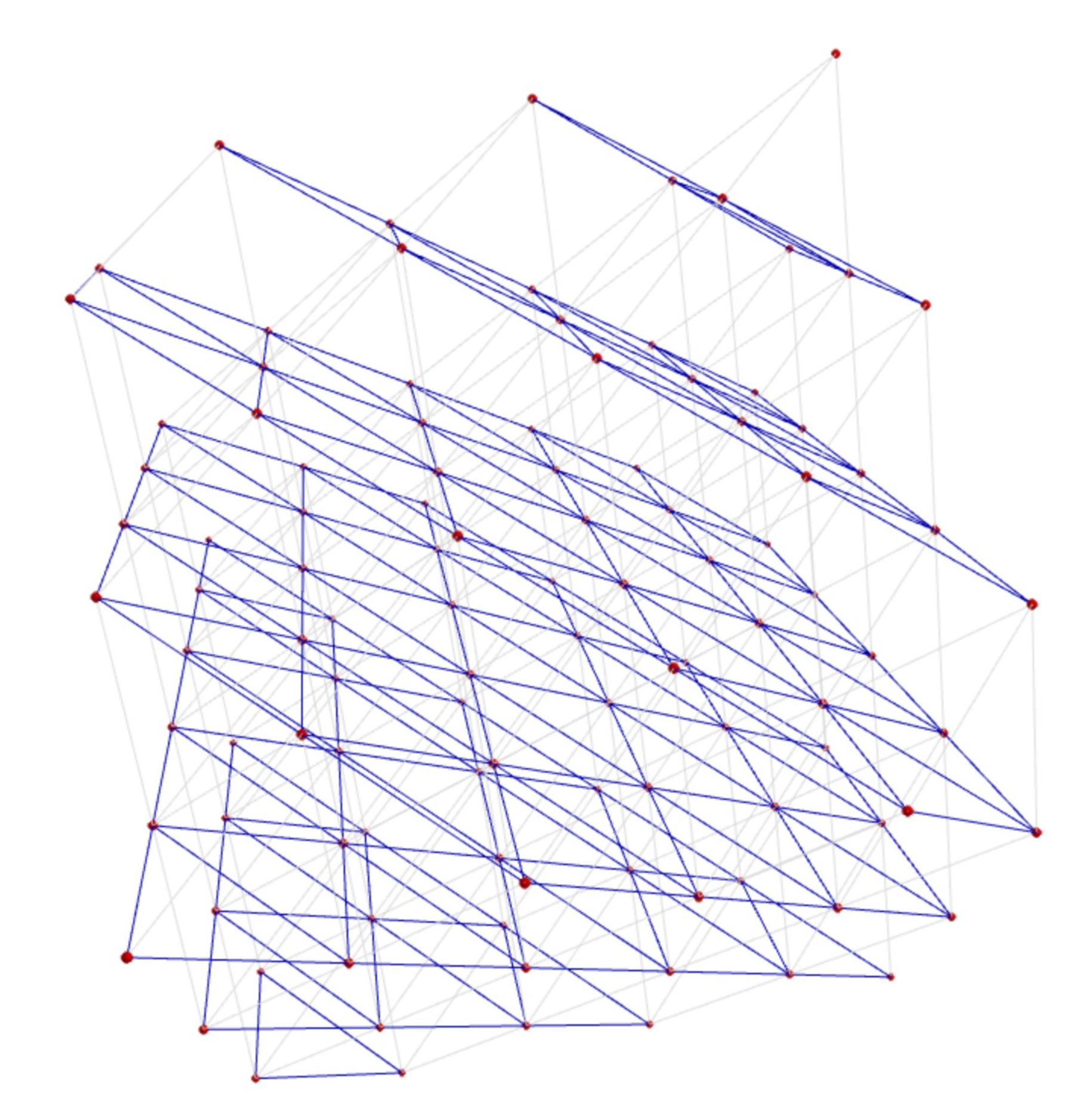}
    \caption{Syndrome graph for the 6-ring fusion network with $ZZ$ measurement outcomes highlighted in blue and $XX$ measurement outcomes in grey. The red nodes are checks. The $ZZ$ measurement outcomes form disconnected plains of the triangular lattice.}
    \label{fig:structured_failure}
\end{figure}

However, using static bias arrangement, it is possible to choose the failure bases in a way that allows the fusion network to operate without any additional encoding or boosting. In the syndrome graph for the 6-ring fusion network as shown in  Fig.~\ref{fig:kagome6_layout}(e), horizontal and vertical edges correspond to $XX$ measurements and diagonal edges correspond to $ZZ$ measurements from fusions. In Fig.~\ref{fig:structured_failure}, we plot the same syndrome graph with the $ZZ$ measurements highlighted in blue and $XX$ measurements in gray. From this figure, it is easy to see that the $ZZ$ measurements form disconnected 2-dimensional planes, each of which is a triangular lattice.  

The gap between the $25\%$ fusion failure probability of boosted fusion and $34.7\%$ maximum tolerable erasure means that there is room to tolerate photon loss and measurement error. In the left of Fig.~\ref{fig:K6_threshold_curve}, we plot the logical error rate for three code sizes. The crossing of these three curves gives us the loss threshold which is $0.48\%$. Further details of the numerical simulations are presented in section ~\ref{subsubsec:numerical_results}.

While static bias arrangement allows us to tolerate fusion failure, it results in a relatively low tolerance to loss. This is because even a small loss rate results in erasure of $XX$ measurements which then links the triangular planes. The tolerance to fusion failure comes from reducing the dimensionality of the erasures and these links between planes remove this advantage very quickly. In the high loss regime, deciding the failure basis only based on erasures from fusion failure is insufficient; the decision needs to be based on erasures from photon loss as well. We describe such a method in the next section.

\section{Dynamic bias arrangement}

In this section we describe a method for choosing the failure basis based on previously observed erasures. This is a general method that can be used to increase the loss tolerance of any fusion network. We study the particular case of the 6-ring fusion network where we find that the method more than triples the loss tolerance. 

\begin{figure}[h]
    \centering
    \includegraphics[width=0.7\columnwidth]{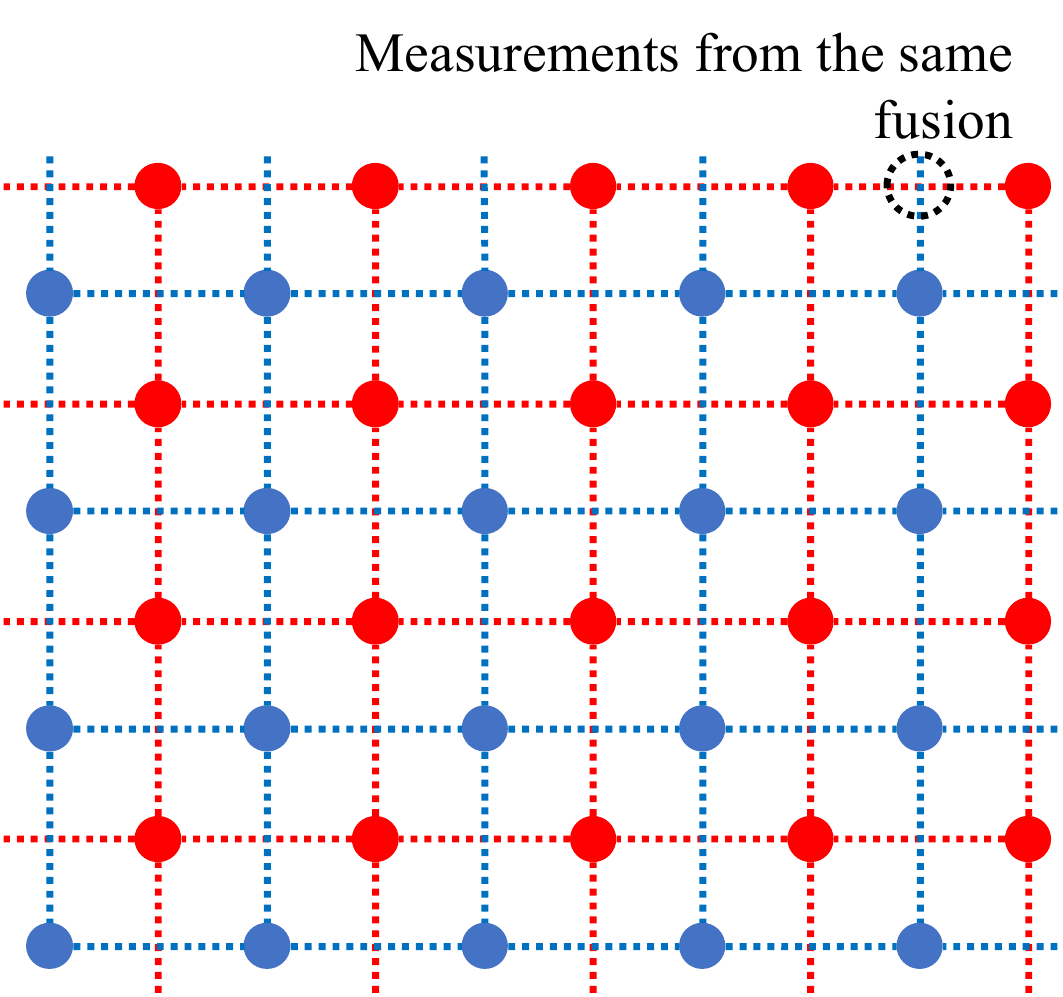}
    \caption{Square Primal (blue) and dual (red) syndrome graphs for an example fusion network. Edges sharing the same midpoint correspond to measurements from the same fusion. Based on the choice of fusion basis, the erasure probability of one can be suppressed at the cost of the other. The square syndrome graph here is only used for pedagogical ease. The numerical results are based on the syndrome graphs for the 6-ring fusion network. In the following figures, it will also be easier to separate the primal and dual syndrome graphs. 
    }
    \label{fig:square_primal_dual_mapping}
\end{figure}

To introduce the methods of dynamic bias arrangement we will first describe them using a 2-d illustrative model of a square syndrome graph as depicted in Fig.~\ref{fig:square_primal_dual_mapping} for pedagogical ease. These methods can be directly applied to any syndrome graph, and the numerical results we present later are for the 3-dimensional structure of the 6-ring fusion network.

When a fusion is made, we get two outcomes that correspond to two edges in the syndrome graph, one in the primal and one in the dual. At a given point in time the system will contain both measured and unmeasured fusions.
The state of the system can be represented by labelling edges in the syndrome graph. We will consider three types of edges in a syndrome graph: 
\begin{enumerate}
    \item \textbf{Erased edges} represent measurements that have been attempted, and where the outcomes were erased. We will depict these as black solid lines,
    \item \textbf{Measured edges} represent measurements that have been successfully made. We will depict these as green solid lines.
    \item \textbf{Unmeasured edges}, that are measurements that have not been performed yet and could be erased or measured once they are performed.
\end{enumerate}

One of the features of photonic quantum computing is that hardware can be re-used in time to create many physical qubits contributing to the same logical qubit using the technique of interleaving as described in \cite{bombin2021interleaving}. In this architecture the measurements in FBQC can be performed sequentially (as opposed to in layers, or all at once). Here we consider a scenario where fusions are performed in a sequential manner.

\subsubsection{Connected components} 

We define a \textit{connected component} in a syndrome graph as the set of nodes (parity checks) that are connected via erased edges. 
If there is a connected component that spans the logical qubit in either the primal or dual syndrome graph, as shown in Fig.~\ref{fig:logical_erasure_vs_not}(a), then the logical qubit sees a logical erasure. However, if the same number of erasures are arranged in a way that they do not span the logical qubit, as shown in Fig.~\ref{fig:logical_erasure_vs_not}(b), the logical qubit does not see any error. If every edge were erased in an i.i.d fashion in a large lattice, the presence of a spanning path (i.e. logical error) and a connected component of the same order as the full lattice have the same threshold which is equal to the bond percolation threshold of the syndrome graph~\cite{barrett2010fault}.

\subsection{Principles of Exposure Based Adaptivity (XBA)}

\begin{figure}[h]
    \centering
    \includegraphics[width=\columnwidth]{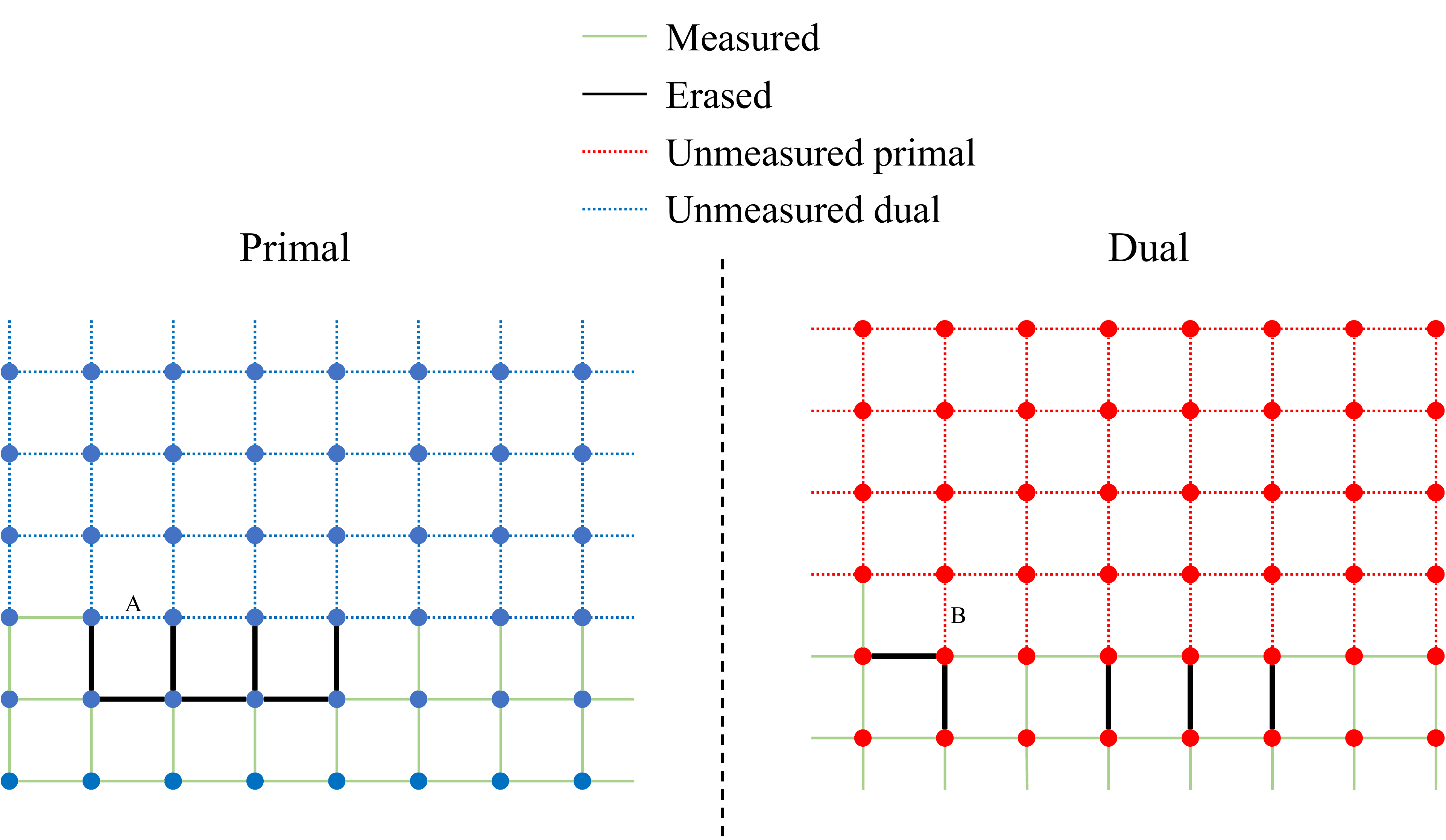}
    \caption{The ``trivial" dynamic bias arrangement decision. The fusion is attempting to measure edges A and B. Since A is a useless intra-cluster edge, the fusion basis is chosen so that A is erased on fusion failure, which reduces the erasure probability for B, which could still be useful.}
    \label{fig:intracluster_decision}
\end{figure}

We will now describe a heuristic for choosing the failure basis based on previous erasures that we call ``exposure based adaptivity" or XBA.
The XBA heuristic attempts to prevent the growth of large connected components. Consider a fusion that measures two syndrome graph edges, A and B. If the two nodes adjacent to one of the edges are part of the same connected components, we call the edge an \textit{intra-cluster edge} as in the case of edge A in the Primal (left) syndrome graph in Fig.~\ref{fig:intracluster_decision} . The measurement corresponding to an intra-cluster edge is useless for fault-tolerance in this setting\footnote{One exception is when an edge completes a loop traversing a logical qubit with periodic boundary conditions. However this is rare in the below threshold regime. We use the same decision-making rule for all intra-cluster edges for convenience.}. It cannot contribute to any useful parity check. In terms of the syndrome graph, it cannot connect nodes that are not already connected. In such a situation, we choose the failure basis so that the measurement probability of A is sacrificed which allows us to reduce the erasure probability from the other measurement from the fusion, labelled as B in the dual, which could still be useful. This is a ``trivial" dynamic bias arrangement decision because only one of the measurements from the fusion is useful, and therefore the prioritization is clear.

\begin{figure}[h]
    \centering
    \includegraphics[width=\columnwidth]{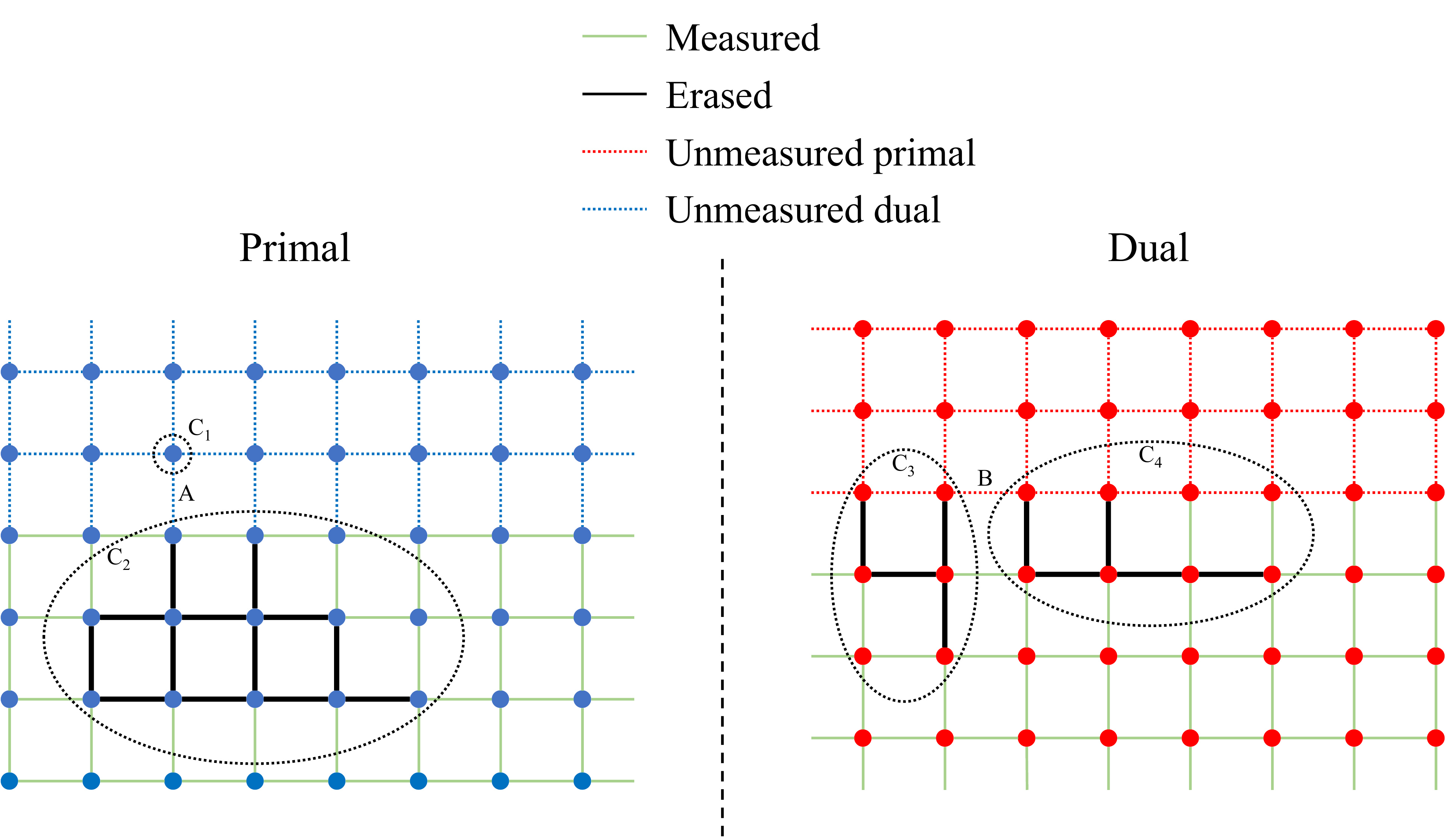}
    \caption{An example of a non-trivial dynamic bias arrangement decision. The fusion being performed measures edges $A$ and $B$. The adjacent clusters and their exposures are $C1: 3$, $C_2: 1$, $C_3: 4$, $C_4: 4$. As a result, the exposures of the edges are $A: 3$, $B: 16$. Since the exposure of $B$ is greater, the fusion basis is chosen such that $B$ is measured and $A$ is erased in case of fusion failure. In this example, accretion $a = 0$.}
    \label{fig:large_lowexposure_example}
\end{figure}

If neither measurement from a fusion is useless, we need to judge which edge is more important for preventing logical error. Although the size of the connected components adjacent to an edge seems to be a good heuristic for determining the importance of an edge, this is not always true. Consider the example in the primal syndrome graph of Fig.~\ref{fig:large_lowexposure_example}. Although there is a large cluster $C_2$ (inside the dotted oval) adjacent to edge $A$ in the primal graph, most of the edges adjacent to it have been successfully measured. As a result, the connected component has very limited potential for growth. Such a situation where most of the edges adjacent to a connected component are successful is quite likely in an fusion based quantum computer where resource state generators are re-used since fusions are performed sequentially~\cite{bombin2021interleaving}.

\subsection{Defining Exposure}

In order to make the decision on the relative importance of the two fusion measurement outcomes we introduce a quantity called \emph{exposure}. From this quantity we can develop a heuristic decision making strategy to implement the ideas above. Exposure can be defined for a connected component, or an edge as follows:  

\begin{definition}
The exposure of a connected component is calculated by summing over all the edges adjacent to the nodes of the cluster. Each unmeasured edge contributes a value 1 to the sum, and each measured edge contributes a value $a<1$.
\end{definition}

\begin{definition}
The exposure of an unmeasured edge $E$ is defined as the product of the exposures of the components at its two ends, where the edge $E$ itself is excluded from the calculation of exposure of the adjacent clusters. 
\end{definition}

The exposure of a connected component is an indicator of its potential for growth, where $a$ is a variable parameter called \textit{accretion}. In the presence of Pauli errors, a large number of measured edges adjacent to a large erasure cluster are undesirable and we set $a>0$ to account for this.

The exposure of an edge takes into account both of the clusters it would connect to if erased. The choice of using the product of exposures at the ends of an edge is inspired by the problem of explosive percolation that has been studied in graph theory as a way of delaying percolation~\cite{achlioptas2009explosive}. 

\subsection{Determining the fusion choice}

Given multiple ways in which a fusion can be biased, we now describe how the fusion choice is determined.

Consider a fusion that measures two syndrome graph edges $E_1$ and $E_2$ with exposures $\chi_1$ and $\chi_2$. Based on $\chi_1$, $\chi_2$ and information on whether $E_1$ and/or $E_2$ are intra-cluster edges, we quantify the relative importance of the two edges with a quantity we call bias. The bias lies in the interval $[0,1]$. A value closer to $0$ indicates that the $E_1$ is more important and a value closer to $1$ indicates $E_2$ is more important. If both measurements from the fusion are equally important, bias $1/2$.

To compute the bias, we first define a quantity intermediate bias (IB) as 

\begin{equation}
\text{IB} = \begin{cases} 
      \frac{1}{2}\left(\frac{\chi_{2}}{\chi_{1}}\right)^{\xi} & \chi_{1}\geq \chi_{2} \\
      1 - \frac{1}{2}\left(\frac{\chi_{1}}{\chi_{2}}\right)^{\xi} & \chi_{1}< \chi_{2} 
   \end{cases}
\end{equation}

where $\xi$ is a parameter that we call the exposure exponent. Setting $\xi$ to a small value pushes the intermediate bias (IB) towards 1/2 i.e. it tends to push us towards not being too biased towards one parity. The exposure exponent for any nonzero value does not limit the range in which IB can vary.

We also define a parameter \textit{squeeze}, $s$, which limits the range in which the final bias can vary if the fusion is not ``trivial". A fusion is trivial when at least one of the edges is an intra-cluster edge.

Our final bias is defined as 

\begin{equation}
\text{bias} = \begin{cases}
        0 & E_2\text{ is intra-cluster} \\
      1 & E_1\text{ is intra-cluster}\\
      (1-s)\text{IB} + \frac{s}{2} & \text{non-trivial fusion}.\\
   \end{cases}
\end{equation}
   
Squeeze closes the window in which bias can vary around 1/2 and avoids extreme biasing decisions in non-trivial cases. In the case of $s=1$, the bias of all non-trivial fusions is 1/2 and with $s=0$, the bias of non-trivial fusions has the maximum range of $[0, 1]$. For trivial fusions, the bias is either $0$ or $1$. The bias doesn't matter if both $E_1$ and $E_2$ are intra-cluster and in such cases, the bias is set to 0.

An example with the trivial rule was shown in Fig.~\ref{fig:intracluster_decision}. Fig.~\ref{fig:large_lowexposure_example} shows an example with a non-trivial decision. The fusion being performed measures edges $A$ and $B$. If $a=0$, the adjacent clusters and their exposures are $C1: 3$, $C_2: 1$, $C_3: 4$, $C_4: 4$. As a results the exposures of the edges are $A: 3$, $B: 16$. Since the exposure of $B$ is greater, the bias is chosen such that the probability of $B$ being erased is less than $A$. The parameters $a$, $\xi$ and $s$ can be optimized to tailor the decision making to the fusion network. 

Given multiple ways in which the fusion can be biased, we pick the fusion option that minimizes the following cost function:
\begin{equation}
 (1-\text{bias})\times \text{Prob}(E_1 \text{ is erased}) + \text{bias}\times \text{Prob}(E_2 \text{ is erased}).
 \label{eq:bias_objective}
\end{equation}

The number of options over which Eq.~\ref{eq:bias_objective} is minimized will depend on the type of fusion. In the unencoded physical fusions that we first consider, there are only four options coming from the choice of boosting and failure basis. The case of encoded fusion as described in Section ~\ref{subsec:encoded_fusion} has more options.

\subsection{Numerical results}
\label{subsubsec:numerical_results}

We perform numerical simulations of the error correction procedure by performing Monte Carlo sampling of the error distribution with dynamic bias arrangement, and evaluating the effect of decoding. We consider the 6-ring fusion network and error model specified previously with periodic boundary conditions in all three dimensions, with system sizes $L=12,16,20$. We sweep over a series of error parameters and for each parameter combination we perform 100,000 individual numerical experiments and combine the results to compute a logical error rate. The logical error rate is defined as the probability of any logical error in any logical qubit encoded by the fusion network.

In order to sample errors, each fusion is sampled sequentially in an order that corresponds to a rastered interleaving scheme~\cite{bombin2021interleaving}. As shown in Fig.~\ref{fig:kagome6_layout}, resource states in the 6-ring network are positioned along the vertices of a cubic lattice with one fusion in each of the six cartesian directions $x+$, $x-$, $y+$, $y-$, $z+$ and $z-$. Resource states are added along the $x$ axis followed by the $y$ axis followed by the $z$ axis, such that the resource state at site ($x,y,z$) is resource state number $(x+yL+zL^2)$ in the order that it is added to the bulk. Whenever a new resource state is added to the bulk, neighboring fusions in the directions $x-$, $y-$ and $z-$ are performed sequentially in that order. At each step the XBA heuristic described earlier is applied to decide on the fusion configuration based on the location of erasures in previous steps, and the outcomes of the fusion are then sampled according according to the relevant error model. 

Here we only consider the limiting case where all fusions are performed sequentially. While this is optimal for the threshold, XBA can also work with less information if desired in a physical implementation; edges whose state is unknown can be marked as unmeasured. The decision-making algorithm only requires information about neighboring clusters. Cluster exposure can be tracked using a disjoint sets data structure~\cite{galler1964improved, tarjan1975efficiency, delfosse2021almost}, using which the exposure of each cluster can be tracked efficiently in almost-linear time.

In this paper, we use the parameters $a = 0.5$, $\xi = 0.5$, $s = 0.2$ which were numerically found to perform well across a broad range on loss and measurement error values for the 6-ring network\footnote{Different optimized values of these parameters for different error parameters would lead to slightly better performance but we use the same values in this paper for simplicity.}. 

Since the error model is intrinsically correlated between primal and dual errors, the errors for each fusion outcome in the primal and dual syndrome graph are sampled together. The primal and dual lattices are then decoded independently using a minimum weight perfect matching decoder (MWPM)~\cite{edmonds1965paths, dennis2002topological, kolmogorov2009blossom}. Note that in the loss only case, both the MWPM decoder and union find~\cite{delfosse2021almost} decoder are optimal, it is only in the presence of Pauli errors that there is a small difference in threshold.
We also simulate the same system and error model in the absence of XBA and only with static bias arrangement as described in Section~\ref{sec:SBA}  in order to compare the performance. In this case the time-ordering of the fusions has no effect on the result, and each fusion has a fixed configuration. All other aspects of the simulation are the same. 

We consider four options for each fusion: unboosted fusion with failure in $XX$, unboosted fusion with failure in $ZZ$, boosted fusion with failure in $XX$, and boosted fusion with failure in $ZZ$. As noted previously, the chance of optical loss erasing the more important fusion outcome is larger in a boosted fusion. Thus if the exposure of a given edge from one choice of fusion is very much greater than the other choice then it is preferable to perform an unboosted fusion and further increase the probability of obtaining the more important outcome. In other cases, we boost the fusion to minimize the chance of fusion failure. In general, we pick the option that minimizes the value of Eq.~\ref{eq:bias_objective}.

\begin{figure}[ht!]
    \centering 
    \includegraphics[width = \columnwidth]{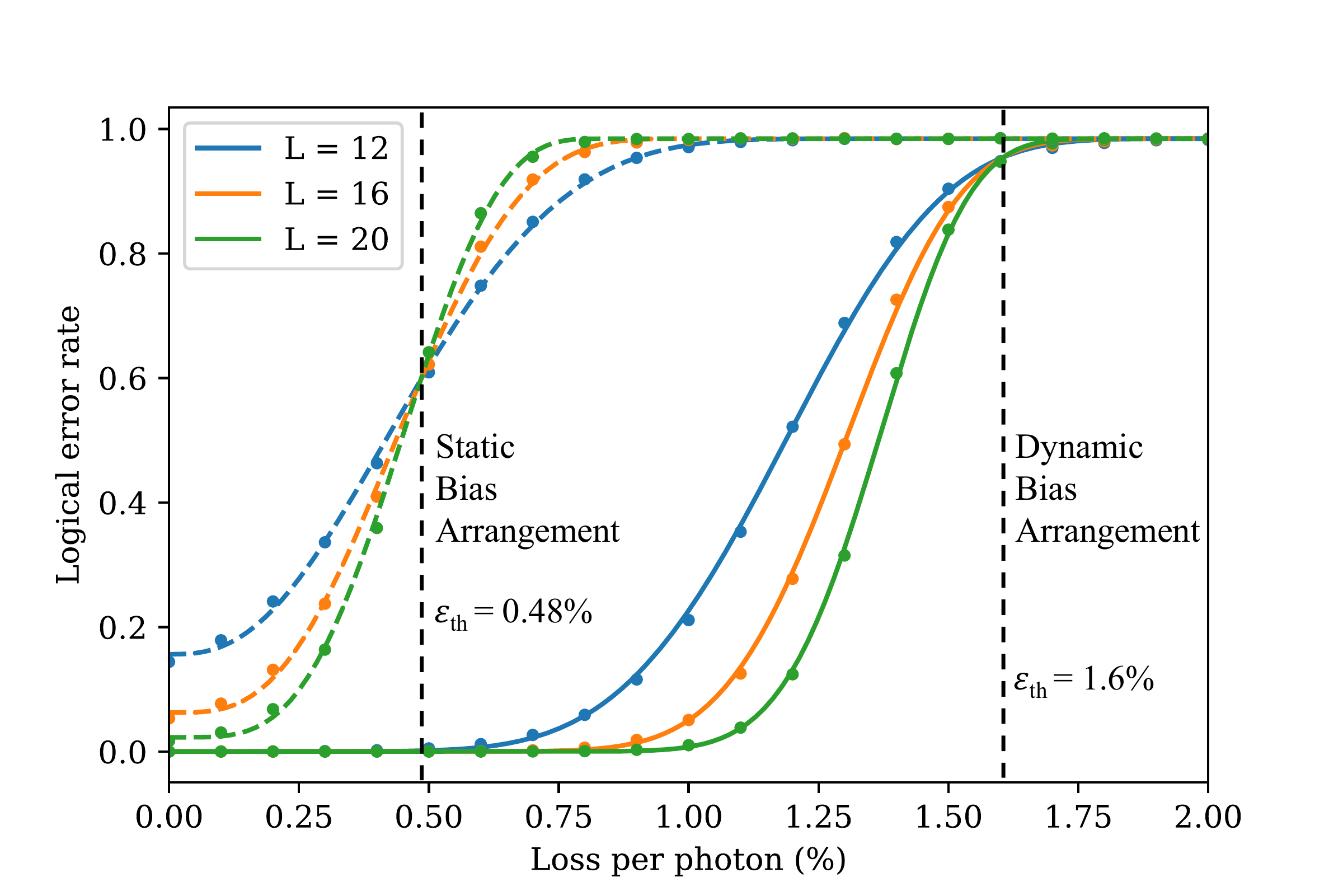}
    \caption{Logical error rate versus photon loss $l$ for the 6-ring network with static bias arrangement (left three curves) and with adaptive arrangement of erasures (right three curves) for different code lengths. The crossing of the curves with different code lengths gives us the loss threshold. Static bias arrangement gives a loss threshold of $0.48\%$. Adaptive arrangement more than triples the loss threshold to $1.6\%$. 
    }
   
    \label{fig:K6_threshold_curve}
\end{figure}

Fig.~\ref{fig:K6_threshold_curve}, shows threshold plots comparing dynamic bias arrangement (solid lines) with static bias arrangement (dashed lines) in the 6-ring fusion network under a loss-only variant of the error model described in section~\ref{subsec:error_model}, where measurement error $p_m = 0$. We find that dynamic bias arrangement more than triples the loss threshold from $0.48\%$ to $1.6\%$. 

\begin{figure}[ht!]
    \centering 
    \includegraphics[width = \columnwidth]{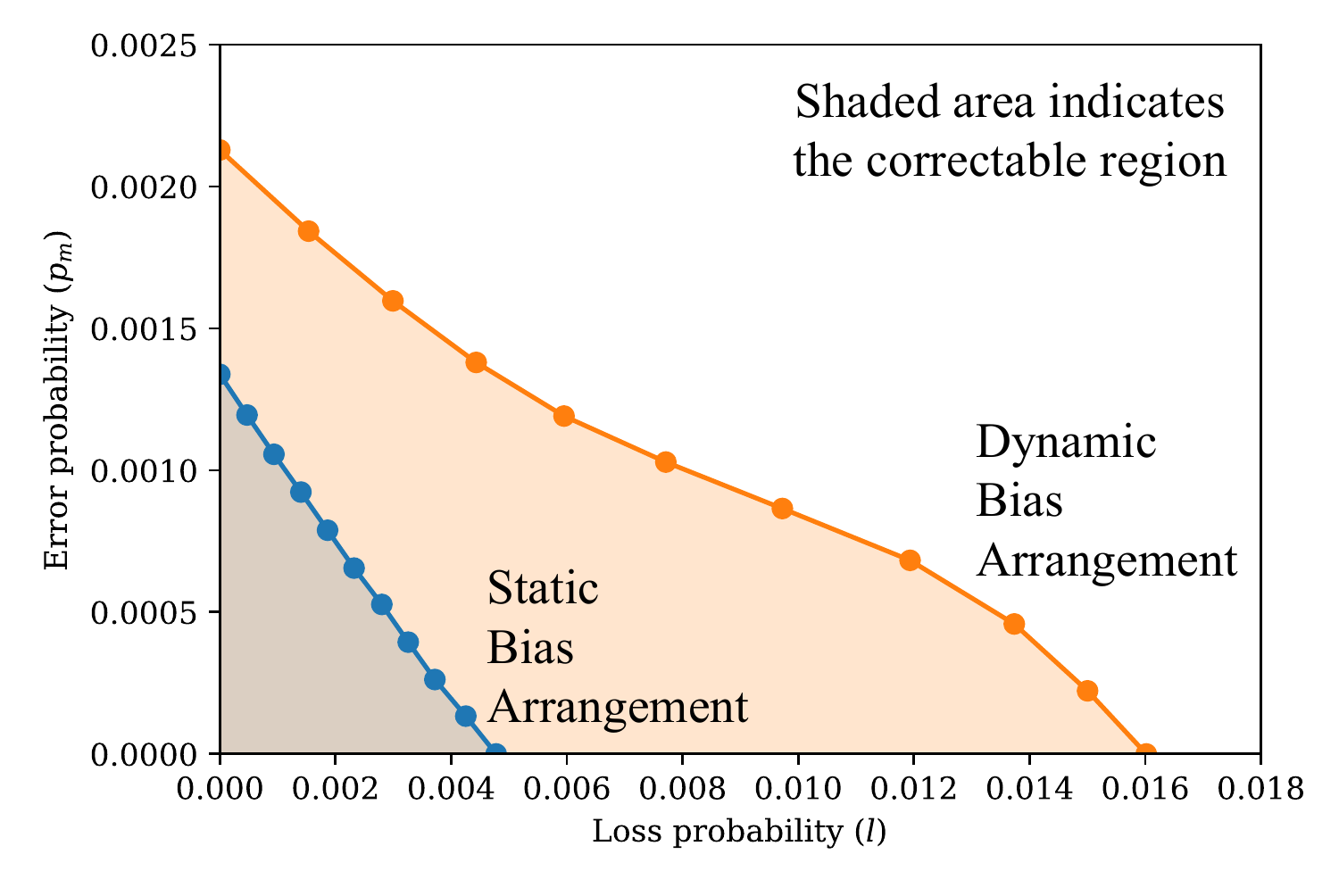}
    \caption{Threshold values of photon loss probability and measurement error probability that can be simultaneously tolerated by the 6-ring network with and without dynamic bias arrangement, plotted for different ratios of loss and measurement error. The threshold with dynamic bias arrangement is significantly better for any ratio of loss and measurement error.}
    \label{fig:Kagome6_phase_diagram}
\end{figure}

Dynamic bias arrangement also gives us an increased threshold in the presence of Pauli errors ($p_m>0$). In Fig.~\ref{fig:Kagome6_phase_diagram} we plot the maximum values of measurement error probability and photon loss that can be simultaneously tolerated by the architecture for different ratios of the two errors. Each point on this plot comes from fitting to a threshold plot for different ratios of measurement error to loss. For all the points plotted here, the threshold is significantly improved with dynamic bias arrangement. The threshold improvement varies between $1.6\times$ in the case of measurement error only and $3.3\times$ in the case of photon loss only. Note that in the case of measurement error only, even though there is no photon loss, there is still significant erasure present from fusion failure, such that dynamic bias arrangement still has a beneficial effect to the threshold.

\subsection{Dynamic bias arrangement with encoded fusion}
\label{subsec:encoded_fusion}

In order to increase loss tolerance, every qubit in the resource state can be encoded in a small local code with code stabilizers $\mathcal{C}$, encoded $X$ operator $\overline{X}$ and encoded $Z$ operator $\overline{Z}$~\cite{bartolucci2021fusion}.

An encoded fusion can be performed between two encoded qubits which we label $A$ and $B$. The code stabilizers and encoded operators for side $A$ are $\mathcal{C}_A$, $\overline{X}_A$ and $\overline{Z}_A$ while those for side $B$ are $\mathcal{C}_B$, $\overline{X}_B$ and $\overline{Z}_B$. The encoded fusion consists of $n$ pairwise physical Bell fusions where $n$ is the number of qubits in the encoding. The fusions are performed sequentially and fusion $i$ attempts to measure $X_{Ai}X_{Bi}$ and $Z_{Ai}Z_{Bi}$. The objective of the encoded fusion is to measure $\overline{X}_A\overline{X}_B$ and $\overline{Z}_A\overline{Z}_B$ by combining measured physical stabilizers which can be done in different ways because of the redundancy introduced by the code stabilizers. This redundancy results in tolerance to loss and fusion failure. 

We assume the encoded fusions are performed sequentially in the same way as the unencoded case. Before each encoded fusion, the bias is calculated. We optimize the fusion measurement basis for every fusion given the bias and previous fusion outcomes in the encoded fusion to minimize the cost function in Eq.~\ref{eq:bias_objective}. Although this optimization is inefficient, it is possible to do this for the small code sizes that we want to use in the resource states. This optimization can be done for different values of bias before any fusion is performed.

As an example, we look at the effect of encoding every resource state qubit in the 6-ring network in the (2,2) Shor code which is a [[4,1,2]] quantum code with code stabilizers $\langle X_1X_2X_3X_4, Z_1Z_2I_3I_4, I_1I_2Z_3Z_4\rangle$ and encoded operators $\overline{X} = X_1X_2I_3I_4$ and $\overline{Z} = Z_1I_2Z_3I_4$. It is easy to see that there are multiple different ways to perform measure the encoded operators from the physical fusion measurements since $\overline{X}_A\overline{X}_B = (X_{A1}X_{B1})(X_{A2}X_{B2}) = (X_{A3}X_{B3})(X_{A4}X_{B4})$ and $\overline{Z}_A\overline{Z}_B = (Z_{A1}Z_{B1})(Z_{A3}X_{B3}) = (Z_{A1}Z_{B1})(Z_{A4}X_{B4}) = (Z_{A2}Z_{B2})(Z_{A3}X_{B3}) = (Z_{A2}Z_{B2})(Z_{A4}X_{B4})$.

\begin{figure}[ht!]
    \centering 
    \includegraphics[width = \columnwidth]{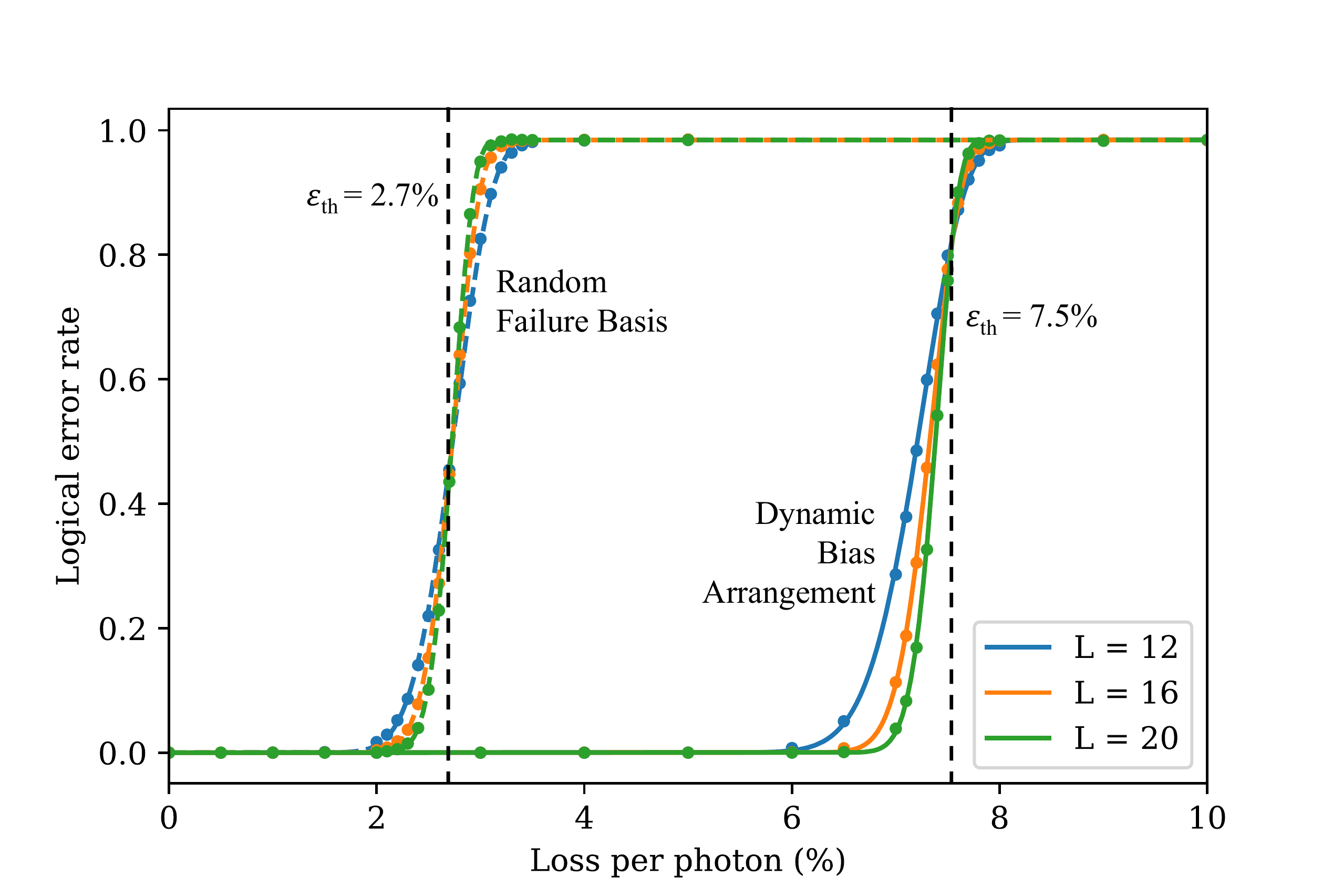}
    \caption{Logical error rate versus photon loss $l$ for the 6-ring network with every qubit encoded with a (2,2) Shor code with dynamic bias arrangement (solid lines). The loss threshold is $7.5\%$ per photon. The loss threshold when the failure basis is chosen randomly~\cite{bartolucci2021fusion} is shown on the left (dashed line) and has a threshold of $2.7\%$ per photon}
    \label{fig:K6_2_2_threshold_curve}
\end{figure}

In Fig.~\ref{fig:K6_2_2_threshold_curve}, we plot the logical error rate for the 6-ring fusion network with every resource state qubit encoded in the (2,2) Shor code where every photon goes through loss $l$. The fusion network can tolerate up to $7.5\%$ loss in every qubit which corresponds to $14.6\%$ loss in an unboosted fusion. This is significantly more than the $2.7\%$ threshold without dynamic bias arrangement of erasures with the same resource state~\cite{bartolucci2021fusion}. At this loss rate, the relative erasure probability of the two measurements coming from the same physical fusions are in the ratio 4.4:1 which shows that dynamic arrangement can work with moderate biases. As opposed to \cite{bartolucci2021fusion}, the physical fusions do not need to be boosted.

\section{Discussion}

The ability to dynamically reconfigure biased operations using  better classical decision making during a fault tolerance protocol can have a very significant impact on tolerance to error. We showed that in a photonic architecture dyanmic bias arrangement can increase the tolerance to photon loss by a factor of three compared to static bias arrangement, with otherwise identical physical operations. We introduced the method of \textit{exposure based adaptivity} or XBA which gives a protocol for biasing fusions based on previous measurement outcomes; the marginal error probabilities of measurements are not reduced, but dynamic biasing allows us to correlate errors in a manner that suppresses the growth of erasure clusters, reducing the logical error rate and increasing the threshold.
This can avoid the need to make more complex quantum resources in order to achieve high thresholds ~\cite{lee2022loss, omkar2022all, bell2022optimising, hilaire2023linear}.

Dynamic bias arrangement is based on the ability to bias measurements in different ways and to make this choice based on previous observations.  We have presented numerical results applying dynamic bias arrangement to a photonic architecture where the decision making is based on erasures in previous measurements, since this is the dominant error type in a photonic system. The bias choice determined the relative erasure rate of the primal and dual syndrome graphs, because this is the natural bias mechanism available. However the same techniques could be applied in other systems with biased operations~\cite{xu2022engineering,puri2020bias,lescanne2020exponential,grimm2019kerr}, and the method can be modified to include syndrome information and Pauli error biasing in systems where the dominant error source and bias mechanisms are different.

In a photonic system, fusions are naturally time-ordered when reusing resource states~\cite{bombin2021interleaving} and the bias can be chosen by the switching network that is used in the preparation of the resource state~\cite{bartolucci2021switch}. Here we have considered the limiting case where all measurements are performed sequentially, but the scheme could also be used with more limited information. Since the choice of bias depends only on the size of neighboring erasure clusters, the scheme can work with information from a more limited radius of information.

The scheme of exposure based adaptivity is a heuristic based on results from percolation theory~\cite{achlioptas2009explosive} but there is no reason to believe that it is optimal. There is room to improve this heuristic and more information (such as syndrome information) could be incorporated in the decision-making process. While XBA can be used in any code with a primal and dual syndrome graph, it may also be possible to further generalize the concept to more general classes of codes \cite{bombin2006topological,panteleev2021quantum,camara2005constructions,tillich2013quantum}.

In the setting of photonic quantum computing we have seen that dynamic bias arrangement can have a very significant impact on loss tolerance. Further developments of this technique may help close the gap between hardware and architectures for large scale fault tolerant quantum computation.

\section*{Author Contributions}

MP and NN first studied fusion failure arrangement in encoded qubits. MP and HB developed the method of static bias arrangement in the 6-ring fusion network, which was independently studied in \cite{bonilla2021xzzx, claes2022tailored, sahay2022tailoring, sahay2023high}. MP developed the XBA method of dynamic bias arrangement. CD wrote the simulation framework. JS developed and ran numerical simulations to identify heuristics for the XBA parameters. MP ran the numerical simulations presented in this paper and MP and NN wrote the manuscript with feedback from CD and JS. HB was on leave during the writing of the manuscript and has not reviewed it in its final form.

\section*{Acknowledgements}

We thank all our colleagues at PsiQuantum for helpful discussions and feedback on the draft, especially Ye-Hua Liu for help with development of the code base, Fernando Pastawski for discussions on local codes, Mercedes Gimeno-Segovia for discussions on system considerations, Sam Roberts and Juan Pablo Bonilla Ataides for discussions on extensions to the scheme, and Terry Rudolph for discussions on photonic circuits for adaptive measurements. We would also like to thank Terry Farrelly, Daniel Litinski, Ryan Mishmash, Karthik Seetharam and David Tuckett for general discussions regarding the ideas in the paper.

\bibliographystyle{unsrt}
\bibliography{references}

\begin{thebibliography}{10}

\bibitem{bonilla2021xzzx}
J~Pablo Bonilla~Ataides, David~K Tuckett, Stephen~D Bartlett, Steven~T Flammia,
  and Benjamin~J Brown.
\newblock The xzzx surface code.
\newblock {\em Nature communications}, 12(1):1--12, 2021.

\bibitem{claes2022tailored}
Jahan Claes, J~Eli Bourassa, and Shruti Puri.
\newblock Tailored cluster states with high threshold under biased noise.
\newblock {\em npj Quantum Information}, 9(1):9, 2023.

\bibitem{sahay2022tailoring}
Kaavya Sahay, Jahan Claes, and Shruti Puri.
\newblock Tailoring fusion-based error correction for high thresholds to biased
  fusion failures.
\newblock {\em arXiv preprint arXiv:2301.00019}, 2022.

\bibitem{sahay2023high}
Kaavya Sahay, Junlan Jin, Jahan Claes, Jeff~D Thompson, and Shruti Puri.
\newblock High threshold codes for neutral atom qubits with biased erasure
  errors.
\newblock {\em arXiv preprint arXiv:2302.03063}, 2023.

\bibitem{bartolucci2021fusion}
Sara Bartolucci, Patrick Birchall, Hector Bombin, Hugo Cable, Chris Dawson,
  Mercedes Gimeno-Segovia, Eric Johnston, Konrad Kieling, Naomi Nickerson,
  Mihir Pant, et~al.
\newblock Fusion-based quantum computation.
\newblock {\em arXiv preprint arXiv:2101.09310}, 2021.

\bibitem{aliferis2008fault}
Panos Aliferis and John Preskill.
\newblock Fault-tolerant quantum computation against biased noise.
\newblock {\em Physical Review A}, 78(5):052331, 2008.

\bibitem{tuckett2018ultrahigh}
David~K Tuckett, Stephen~D Bartlett, and Steven~T Flammia.
\newblock Ultrahigh error threshold for surface codes with biased noise.
\newblock {\em Physical review letters}, 120(5):050505, 2018.

\bibitem{aliferis2009fault}
Panos Aliferis, Frederico Brito, David~P DiVincenzo, John Preskill, Matthias
  Steffen, and Barbara~M Terhal.
\newblock Fault-tolerant computing with biased-noise superconducting qubits: a
  case study.
\newblock {\em New Journal of Physics}, 11(1):013061, 2009.

\bibitem{achlioptas2009explosive}
Dimitris Achlioptas, Raissa~M D'Souza, and Joel Spencer.
\newblock Explosive percolation in random networks.
\newblock {\em science}, 323(5920):1453--1455, 2009.

\bibitem{rudolph2017optimistic}
Terry Rudolph.
\newblock Why i am optimistic about the silicon-photonic route to quantum
  computing.
\newblock {\em APL photonics}, 2(3):030901, 2017.

\bibitem{bartolucci2021switch}
Sara Bartolucci, Patrick Birchall, Damien Bonneau, Hugo Cable, Mercedes
  Gimeno-Segovia, Konrad Kieling, Naomi Nickerson, Terry Rudolph, and Chris
  Sparrow.
\newblock Switch networks for photonic fusion-based quantum computing.
\newblock {\em arXiv preprint arXiv:2109.13760}, 2021.

\bibitem{bombin2021interleaving}
Hector Bombin, Isaac~H Kim, Daniel Litinski, Naomi Nickerson, Mihir Pant,
  Fernando Pastawski, Sam Roberts, and Terry Rudolph.
\newblock Interleaving: Modular architectures for fault-tolerant photonic
  quantum computing.
\newblock {\em arXiv preprint arXiv:2103.08612}, 2021.

\bibitem{hein2004multiparty}
Marc Hein, Jens Eisert, and Hans~J Briegel.
\newblock Multiparty entanglement in graph states.
\newblock {\em Physical Review A}, 69(6):062311, 2004.

\bibitem{browne2005resource}
Daniel~E Browne and Terry Rudolph.
\newblock Resource-efficient linear optical quantum computation.
\newblock {\em Physical Review Letters}, 95(1):010501, 2005.

\bibitem{puri2020bias}
Shruti Puri, Lucas St-Jean, Jonathan~A Gross, Alexander Grimm, Nicholas~E
  Frattini, Pavithran~S Iyer, Anirudh Krishna, Steven Touzard, Liang Jiang,
  Alexandre Blais, et~al.
\newblock Bias-preserving gates with stabilized cat qubits.
\newblock {\em Science advances}, 6(34):eaay5901, 2020.

\bibitem{lescanne2020exponential}
Rapha{\"e}l Lescanne, Marius Villiers, Th{\'e}au Peronnin, Alain Sarlette,
  Matthieu Delbecq, Benjamin Huard, Takis Kontos, Mazyar Mirrahimi, and Zaki
  Leghtas.
\newblock Exponential suppression of bit-flips in a qubit encoded in an
  oscillator.
\newblock {\em Nature Physics}, 16(5):509--513, 2020.

\bibitem{xu2022engineering}
Qian Xu, Joseph~K Iverson, Fernando~GSL Brand{\~a}o, and Liang Jiang.
\newblock Engineering fast bias-preserving gates on stabilized cat qubits.
\newblock {\em Physical Review Research}, 4(1):013082, 2022.

\bibitem{barrett2010fault}
Sean~D Barrett and Thomas~M Stace.
\newblock Fault tolerant quantum computation with very high threshold for loss
  errors.
\newblock {\em Physical review letters}, 105(20):200502, 2010.

\bibitem{galler1964improved}
Bernard~A Galler and Michael~J Fisher.
\newblock An improved equivalence algorithm.
\newblock {\em Communications of the ACM}, 7(5):301--303, 1964.

\bibitem{tarjan1975efficiency}
Robert~Endre Tarjan.
\newblock Efficiency of a good but not linear set union algorithm.
\newblock {\em Journal of the ACM (JACM)}, 22(2):215--225, 1975.

\bibitem{delfosse2021almost}
Nicolas Delfosse and Naomi~H Nickerson.
\newblock Almost-linear time decoding algorithm for topological codes.
\newblock {\em Quantum}, 5:595, 2021.

\bibitem{edmonds1965paths}
Jack Edmonds.
\newblock Paths, trees, and flowers.
\newblock {\em Canadian Journal of mathematics}, 17:449--467, 1965.

\bibitem{dennis2002topological}
Eric Dennis, Alexei Kitaev, Andrew Landahl, and John Preskill.
\newblock Topological quantum memory.
\newblock {\em Journal of Mathematical Physics}, 43(9):4452--4505, 2002.

\bibitem{kolmogorov2009blossom}
Vladimir Kolmogorov.
\newblock Blossom v: a new implementation of a minimum cost perfect matching
  algorithm.
\newblock {\em Mathematical Programming Computation}, 1:43--67, 2009.

\bibitem{lee2022loss}
Seok-Hyung Lee, Srikrishna Omkar, Yong~Siah Teo, and Hyunseok Jeong.
\newblock Loss-tolerant linear optical quantum computing under nonideal fusions
  using multiphoton qubits.
\newblock {\em arXiv preprint arXiv:2207.06805}, 2022.

\bibitem{omkar2022all}
Srikrishna Omkar, Seok-Hyung Lee, Yong~Siah Teo, Seung-Woo Lee, and Hyunseok
  Jeong.
\newblock All-photonic architecture for scalable quantum computing with
  greenberger-horne-zeilinger states.
\newblock {\em PRX Quantum}, 3(3):030309, 2022.

\bibitem{bell2022optimising}
Tom~J Bell, Love~A Pettersson, and Stefano Paesani.
\newblock Optimising graph codes for measurement-based loss tolerance.
\newblock {\em arXiv preprint arXiv:2212.04834}, 2022.

\bibitem{hilaire2023linear}
Paul Hilaire, Yaron Castor, Edwin Barnes, Sophia~E Economou, and
  Fr{\'e}d{\'e}ric Grosshans.
\newblock Linear optical logical bell state measurements with optimal
  loss-tolerance threshold.
\newblock {\em arXiv preprint arXiv:2302.07908}, 2023.

\bibitem{grimm2019kerr}
Alexander Grimm, Nicholas~E Frattini, Shruti Puri, Shantanu~O Mundhada, Steven
  Touzard, Mazyar Mirrahimi, Steven~M Girvin, Shyam Shankar, and Michel~H
  Devoret.
\newblock The kerr-cat qubit: stabilization, readout, and gates.
\newblock {\em arXiv preprint arXiv:1907.12131}, 2019.

\bibitem{bombin2006topological}
Hector Bombin and Miguel~Angel Martin-Delgado.
\newblock Topological quantum distillation.
\newblock {\em Physical review letters}, 97(18):180501, 2006.

\bibitem{panteleev2021quantum}
Pavel Panteleev and Gleb Kalachev.
\newblock Quantum ldpc codes with almost linear minimum distance.
\newblock {\em IEEE Transactions on Information Theory}, 68(1):213--229, 2021.

\bibitem{camara2005constructions}
Thomas Camara, Harold Ollivier, and J-P Tillich.
\newblock Constructions and performance of classes of quantum ldpc codes.
\newblock {\em arXiv preprint quant-ph/0502086}, 2005.

\bibitem{tillich2013quantum}
Jean-Pierre Tillich and Gilles Z{\'e}mor.
\newblock Quantum ldpc codes with positive rate and minimum distance
  proportional to the square root of the blocklength.
\newblock {\em IEEE Transactions on Information Theory}, 60(2):1193--1202,
  2013.

\end{thebibliography}

\end{document}